# Tailorable stimulated Brillouin scattering in nanoscale silicon waveguides.

Heedeuk Shin[1], Wenjun Qiu[2], Robert Jarecki[1], Jonathan A. Cox[1], Roy H. Olsson III[1], Andrew Starbuck[1], Zheng Wang[3], and Peter T. Rakich[1,4*]

[1]Sandia National Laboratories, PO Box 5800 Albuquerque, NM 87185 USA
[2]Department of Physics, Massachusetts Institute of Technology, Cambridge, MA 02139 USA
[3]Department of Electrical and Computer Engineering, University of Texas at Austin, Austin, TX 78758 USA
[4]Department of Applied Physics, Yale University, New Haven, CT 06520 USA
[*]email: rakich@alum.mit.edu

**Photon-phonon coupling through travelling-wave stimulated Brillouin scattering provides a powerful means of realizing tailorable slow light states [1–3], RF-photonic signal processing[4–6], narrow-line-width laser sources[7–11], RF-waveform synthesis[12–15], and optical frequency comb-generation[12,16,17] in guided-wave systems. Silicon-based CMOS compatible technologies of this type could enable high performance signal processing applications through nano-scale Brillouin interactions[18]. Nanoscale modal confinement has been shown to radically enhance nonlinear light-matter interactions within silicon waveguides[19–23] and cavity optomechanics[24–30]. However, traveling-wave Brillouin nonlinearities have yet to be observed in silicon nanophotonics. Through a new class of hybrid photonic-phononic waveguides utilizing phononic mode engineering, we demonstrate tailorable travelling-wave forward-stimulated Brillouin scattering nonlinearities in nanophotonic silicon waveguides for the first time. The resulting Brillouin nonlinearities are 3000 times stronger than forward-stimulated Brillouin response than previously reported waveguide systems[12,31]. Experiments reveal that a coherent combination of material-induced electrostrictive forces and boundary-induced radiation pressure greatly enhance Brillouin nonlinearities at**

**new form of boundary-induced Brillouin nonlinearity[32–34], and a new regime of boundary-mediated Brillouin coupling. Structural tuning of phononic resonances from 1-18 GHz with high quality-factor ( > 1000 ) results in tailorable nonlinear optical susceptibilities due to the coherent interference of Kerr and Brillouin effects. Moreover, high frequency and wide-band coherent phonon emission, produced through such Brillouin processes, paves the way towards seamless integration of chip-scale silicon photonic, MEMS, and CMOS signal processing technologies.**

Silicon-based cavity-optomechanical systems have recently enabled powerful new forms of quantum state transfer[25,35], slow light[36], phonon lasers[37], and optomechanical ground-state cooling[38]. Such cavity systems exploit resonantly enhanced coupling between discrete photonic and phononic modes. As a fundamental complement to cavity systems, traveling-wave Brillouin processes produce coupling between a continuum of photon and phonon modes for a host of wideband (0.1-34 GHz) radio frequency (RF) and photonic signal processing applications[12,34,39–41]. For example, traveling-wave Brillouin processes have yielded wide bandwidth pulse compression[13,42], pulse and waveform synthesis[12,14–16,43], coherent frequency comb generation[12,16,17], variable bandwidth optical amplifiers[43,44], reconfigurable filters[45], and coherent beam combining schemes[46]. Such broad bandwidth operations would be challenging for inherently narrow-band cavity optomechanical systems. While there are compelling opportunities for chip-scale technologies, the observation of Brillouin processes in silicon nanophotonics has proven difficult; strong Brillouin nonlinearities require large optical forces and tight confinement of both phonons and photons, conditions that are not met in conventional silicon waveguides[34].

In this paper, we demonstrate traveling-wave forward stimulated Brillouin scattering (forward-SBS) nonlinearities in silicon waveguides for the first time through a novel class of hybrid photonic-phononic waveguides. Confinement of both photons and phonons is achieved using a membrane-suspended waveguide structure, eliminating the substrate pathway for phonon-loss that stifles SBS in conventional silicon-on-insulator waveguides. In contrast to the theoretical work presented in Ref. [34], the compound material device geometry used here yields independent control of photonic and phononic confinement. Separate control of the optical force distribution and phonon mode spectrum enables exceptionally wide-band tailorable Brillouin coupling from 1 to 18 GHz frequencies. Nano-enhanced electrostrictive forces[33] and boundary-induced radiation pressures[32,47] constructively interfere to yield orders of magnitude higher forward-SBS nonlinear coupling than prior systems[32–34]. The emergence of a radiation pressure-induced coupling at nanoscales contrasts with earlier traveling-wave SBS systems, which are mediated solely by electrostrictive material response. Strong light-boundary interactions give rise this new radiation pressure-mediated regime of Brillouin coupling, producing new geometric degrees of freedom through which Brillouin processes can be tailored to an unprecedented degree.

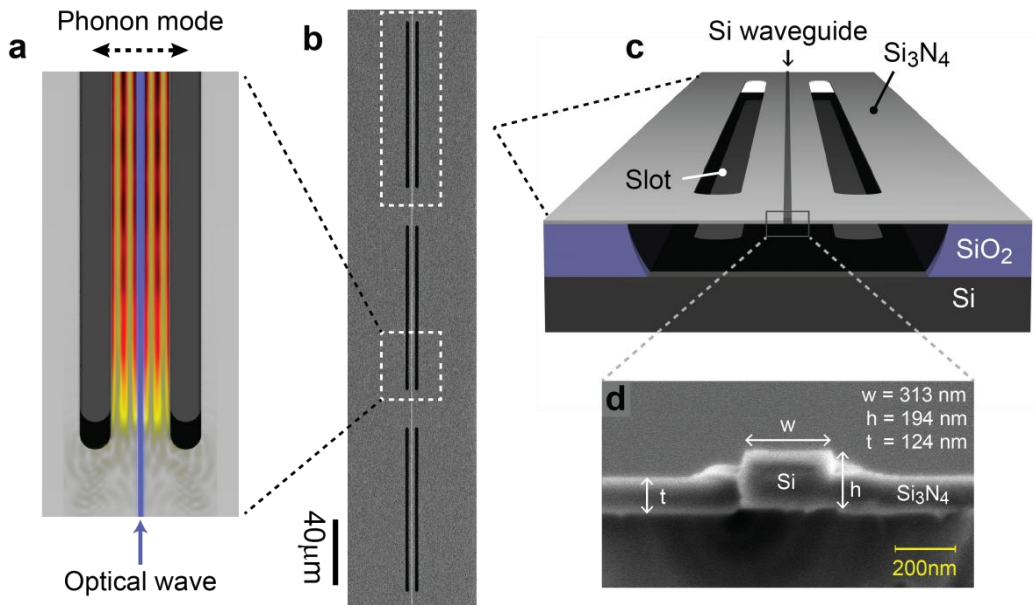

**Figure 1 Hybrid photonic-phononic waveguide enabling independent control of the optical and phonon modes.** **a** Top-down view of a segment of the Brillouin-active waveguide with optical and phonon modes superimposed. Phonon mode displacement is displayed with a red-yellow colour map, while the guided optical mode is in blue. **b** Top-down SEM image of a segment of the transverse phonon-resonator waveguide (3 phononic resonators shown). **c** Schematic showing the anatomy of the hybrid photonic-phononic waveguide system (1 phononic resonator shown). **d** High resolution SEM cross-section of the silicon waveguide core within the nitride membrane.

The hybrid photonic-phononic waveguide under study is seen in the top-down SEM image of Fig. 1b, showing a silicon nanophotonic waveguide embedded within the centre of a series of three suspended phononic resonators. Figure 1c shows the anatomy of the suspended waveguide structure along with an SEM cross-section of the waveguide core (Fig. 1d) consisting of a nanoscale ($313 \times 194$ nm) silicon waveguide embedded in a suspended tensile silicon nitride membrane (thickness 124 nm). This compound material device geometry provides independent control of the photonic and phononic properties of the Brillouin waveguide, enabling the phonon mode spectrum to be tailored independently from the optical force distributions within the core of the silicon waveguide. Total internal reflection between silicon ($n = 3.5$) and silicon nitride ($n = 2.0$) tightly confines the optical mode to the silicon waveguide core, while the patterned silicon nitride membrane acts to confine the generated phonons. As seen from the phase matching diagram of Fig. 2 k-l, the forward-SBS process phase-matches to phonons with a vanishing longitudinal wave-vector (i.e. slow group-velocity guided phonon states). Hence, it is necessary to engineer high Q-factor phononic resonances that support modes of large transverse wave-vector (i.e. perpendicular to the waveguide). This is achieved by truncating the membrane on either side of the silicon waveguide with etched air-slots (of dimension $2 \times 100$ microns) with phononic cavity dimension, $d$, seen in Fig. 1a. The slots efficiently reflect acoustic waves, defining the extended phonon modes of the type seen in Fig 2a. This geometry, which we term the transversely oriented phonon-resonator optical waveguide (TOPROW), produces efficient photon-phonon coupling over a series of discrete phononic resonances between $1 - 18$ GHz, through a traveling-wave forward-SBS process.

The computed optical field distribution for the TE-like guided optical mode of Fig. 2c is shown alongside the waveguide cross-section (Fig. 2b), the electrostrictive force densities (Fig. 2d-f), and radiation pressure-induced force densities (Fig. 2g). Throughout this paper, we explore the optical force mediated photon-phonon coupling between this TE-like optical mode and the phononic modes in Fig. 2a, and Fig. 2h-j. The transverse forces produced within the waveguide core excite an extended phonon mode whose spatial profile within the finite-width suspended waveguide region is seen in Fig. 2a. Full-vectorial multi-physics simulations reveal the elastic wave motion for three characteristic Brillouin-active modes within a suspended section of the waveguide.

Three characteristic Brillouin-active phonon modes with $d = 3.8\,\mu m$ at frequencies 1.28, 3.72, and 6.18 GHz respectively, are shown in Figs. 2h-j corresponding to symmetric Lamb waves.

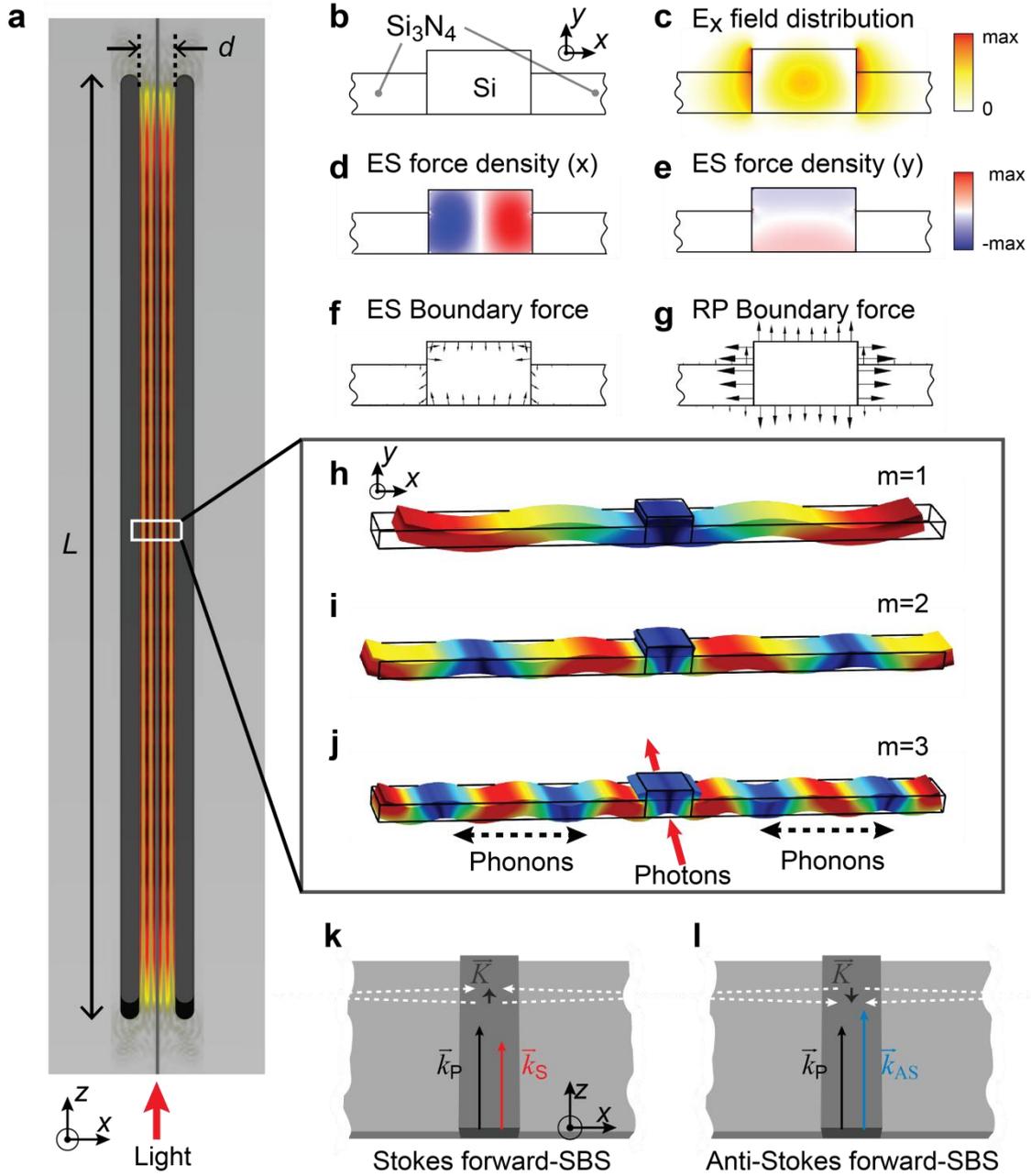

**Figure 2 Simulations of hybrid photonic-phononic waveguide. a** The displacement field of a 3.7 GHz extended phonon excited by optical forces within a TOPROW waveguide with $d = 3.8\,\mu m$ and $L = 100\,\mu m$. **b-c** The waveguide cross-section and the computed $\boldsymbol{E_x}$ field profile of the optical mode. **d-g** Computed force distributions associated with the guided optical mode. **d** and **e** $\boldsymbol{x-}$ and $\boldsymbol{y-}$ components of the electrostrictive (ES) force densities generated within silicon, respectively. **f** and **g** Electrostrictive and radiation pressure (RP) induced boundary forces produced by the optical mode. **h-j** Characteristic phonon modes, with symmetric displacement fields about the waveguide core, which are efficiently coupled to the optical force distributions of **d-g**. **k-l** Phase-matching diagrams

of the respective Stokes and anti-Stokes forward-SBS processes. Phase matching necessitates a vanishing longitudinal phonon wave-vector through forward-SBS, resulting in standing phonon modes (or slow group velocity resonant guided phononic modes) with large transverse wave-vector.

Through experiments, an array of TOPROW devices were studied with phononic cavity dimensions, $d$, between $0.8\,\mu m$ and $3.8\,\mu m$, producing a wide range of Brillouin resonances between 1-18 GHz. In contrast to backward-SBS processes, forward-SBS processes have greatly relaxed phase-matching conditions, enabling efficient excitation of Brillouin active resonances in a wavelength-independent fashion[12,31]. This property of the system allows pump and probe waves, of disparate wavelengths, to couple to each other through the Brillouin response of a single device. To obtain large Brillouin nonlinearities, a serial array of 26 phonon resonators were fabricated along each waveguide, yielding a Brillouin-active length of 2.6 mm out of a total 4.9 mm device length.

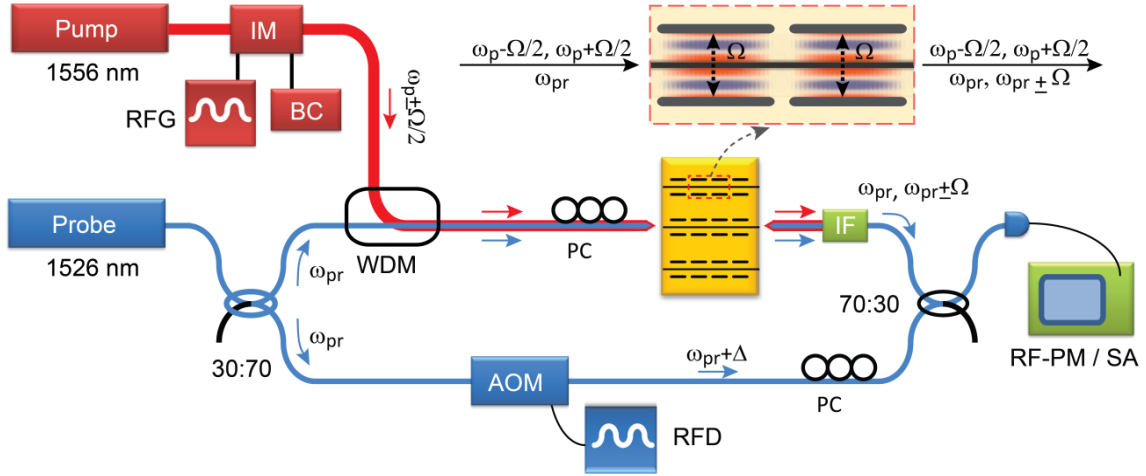

**Figure 3 Heterodyne two-colour pump-probe apparatus used to measure the Brillouin nonlinearity of the TOPROW waveguide.** The nonlinearly induced phase changes imparted to the probe beam (1526 nm) by the waveguide device are detected using heterodyne interferometer, while a modulated pump beam (1556 nm) produces driven excitation of the Brillouin active phonon modes. Components of the apparatus are labelled as follows: ISO is isolator; PC is polarization controller; IM is intensity modulator; RFG is RF generator; BC is bias controller; WDM is wavelength division multiplexer; AOM is acousto-optic modulator; RFD is RF driver; IF is interference filter; SA is spectrum analyser; RF-PM is RF power meter.

Experimental studies of Brillouin nonlinearity were performed with the heterodyne four-wave mixing apparatus, seen in Fig. 3, yielding direct measurement of the third order nonlinear susceptibility. Through four-wave mixing (FWM) experiments, modulated pump (1556 nm) and continuous-wave probe (1526 nm) signals are injected into the device. The modulated pump drives the excitation of Brillouin-active phonons over a wide range of frequencies as the pump modulation frequency is swept. The nonlinear response of the device is then analysed by heterodyne measurement of optical tones imprinted on the disparate probe wavelength due to a coherent combination of the Brillouin and third-order electronic nonlinear susceptibilities (i.e. through four-wave mixing). These sidebands are then analysed as distinct RF tones through heterodyne interferometry, allowing Stokes and anti-Stokes signatures to be resolved separately. For further details, see the Methods Section.

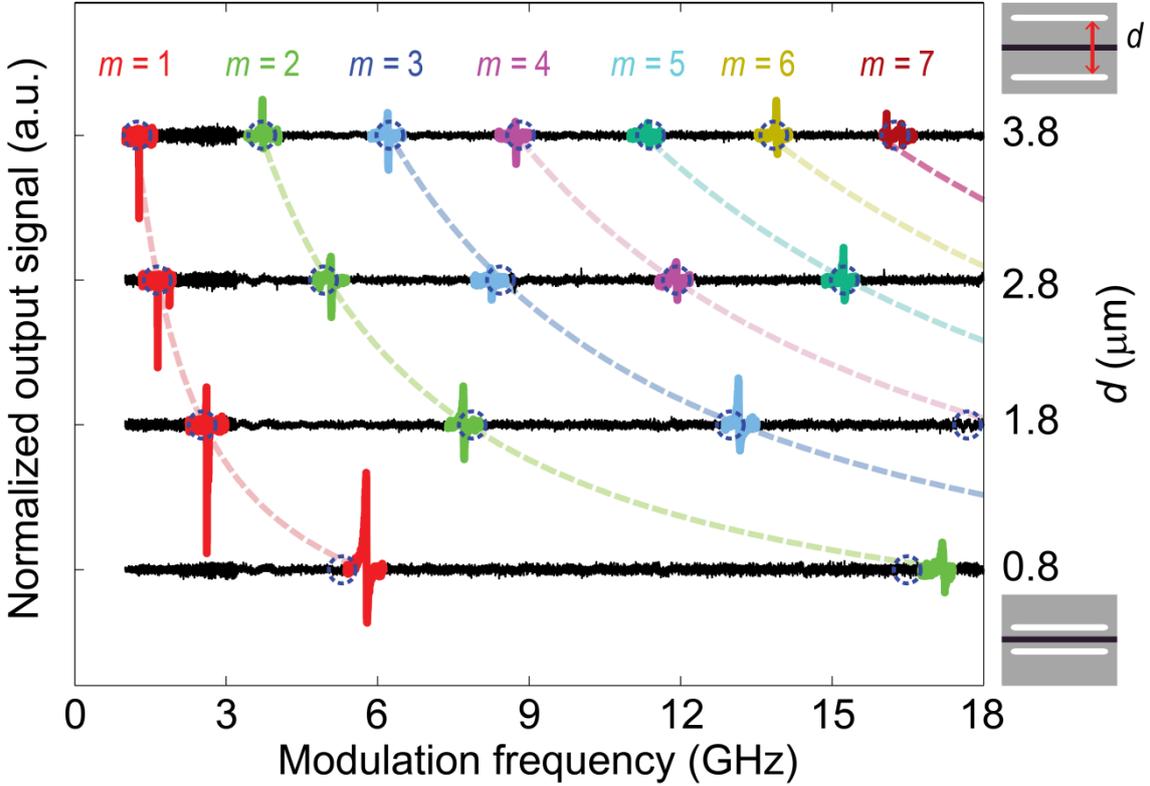

**Figure 4 Spectra of nonlinear Brillouin spectra obtained through heterodyne four-wave mixing measurements.** The resonant Brillouin signatures produced by several TOPROW waveguides are displayed for resonator dimensions of $d = [0.8, 1.8, 2.8, 3.8]$ $\mu m$. Each trace is obtained by normalizing the signal produced by each Brillouin-active waveguide to that of an identical reference silicon waveguide (one that is not Brillouin active) under identical experimental conditions. The dashed blue circles shown atop the experimental traces show the numerically computed frequencies of each Brillouin-active phonon mode as the resonator geometry varies. The upper right inset is a schematic the resonator geometry. Each waveguide has the resonant modes at frequencies of $\omega_0(2m-1)$, where $m = 1,2,3, ...$ is the order of the resonant modes and $\omega_0$ is the fundamental resonant frequency.

As seen from Fig. 4, clear signatures showing the nonlinear Brillouin response were obtained by measuring the intensity of nonlinearly induced sidebands imprinted on the probe as the pump modulation frequency was swept from 1 to 18 GHz. The spectra of Fig. 4 were obtained by integrating the RF power produced by heterodyne detection of the probe signal (including both Stokes and anti-Stokes sidebands) over a discrete set of high-frequency RF bands using RF filters. To remove the unwanted frequency dependence of the detection system and to more clearly exhibit the sharp Brillouin resonances, the spectra of Fig. 4 were normalized to an identical optical waveguide without Brillouin resonators. Analogous to an optical Fabry-Perot cavity, each phonon resonator yields a series of Brillouin resonances with equal frequency spacing. Due to the spatial symmetry of the optical force distribution, only phonon modes with even displacement symmetry with respect to the waveguide core produce efficient Brillouin coupling. The different resonant signatures are colour coded (red, green, blue,...) to indicate the mode order (1$^{st}$, 2$^{nd}$, 3$^{rd}$,...) of each phononic resonance as the Brillouin spectrum shifts with resonator dimension. The frequencies of the experimentally observed resonances show good agreement with simulated mode frequencies over the range of device dimensions. The simulated modal frequencies are displayed as dashed curves atop the experimental data. Figure 4 reveals that a variation of the cavity dimension allows precise placement of Brillouin resonances at virtually any frequency from $1 - 18$ GHz, for an unprecedented degree of nonlinear tailorability. For example, the $m = 2$ resonance (green) is shifted from 3.7 to 18 GHz as the cavity dimension, $d$, is varied from 3.8 to 0.8 $\mu m$. While the bandwidth limitations of our apparatus did not permit measurements beyond 18 GHz, strong Brillouin resonances are expected at 25 GHz and higher frequencies.

Close examination of the resonance signatures in Fig. 4 reveals a Fano-like line-shape produced by each Brillouin resonance, from which the magnitude of the Brillouin nonlinear coefficient, $\gamma_{SBS}$, can be obtained. This line-shape can be more clearly seen from the high resolution spectral scans of Fig. 5a and 5b, which show the line-shape of a characteristic Brillouin resonance ($f = 6.185\ GHz$, with $d = 3.8$) decomposed into its Stokes and anti-Stokes components. These data were obtained by spectrally resolving the distinct heterodyne tones of the Stokes and anti-Stokes signals using a high resolution RF spectrum analyser as the pump modulation frequency was swept. This asymmetric line-shape results from the coherent interference between the Brillouin and electronic Kerr nonlinearities of the waveguide. Involvement of electronic Kerr nonlinearities at the Stokes and anti-Stokes frequencies occurs due to cross-phase modulation (XPM) between the pump and probe beams within the silicon waveguide core. Note that the fibre apparatus yields a negligible contribution to the measured Brillouin and Kerr nonlinearities of the device.

To precisely determine the magnitude of the Brillouin nonlinear coefficient, $\gamma_{SBS}$，relative to the intrinsic Kerr nonlinear coefficient, $\gamma_k$, from these data, the nonlinear coupled amplitude equations were formulated to derive the functional form of the Stokes and anti-Stokes line-shapes (see Supplementary Information). Since SBS is a resonant effect, its nonlinear coefficient takes on a Lorentzian line-shape centred at each Brillouin-active phonon mode. In contrast, the electronic Kerr nonlinearities are non-resonant at 1500 nm wavelengths, yielding a frequency-independent nonlinear coefficient. Analogous to the fibre-based studies of Wang *et. al*, the frequency-dependent interference between the Kerr and Brillouin effects produces the asymmetric (Fano-like) line-shape observed in Fig. 5a and 5b. However, it should be noted that our experimental arrangement is distinct, leading to a different set of coupled amplitude equations. Additionally, nonlinearly generated free carriers in silicon are responsible for the dissimilar line-shapes of the Stokes and anti-Stokes orders, and a larger nonlinear background for frequencies below 2GHz.

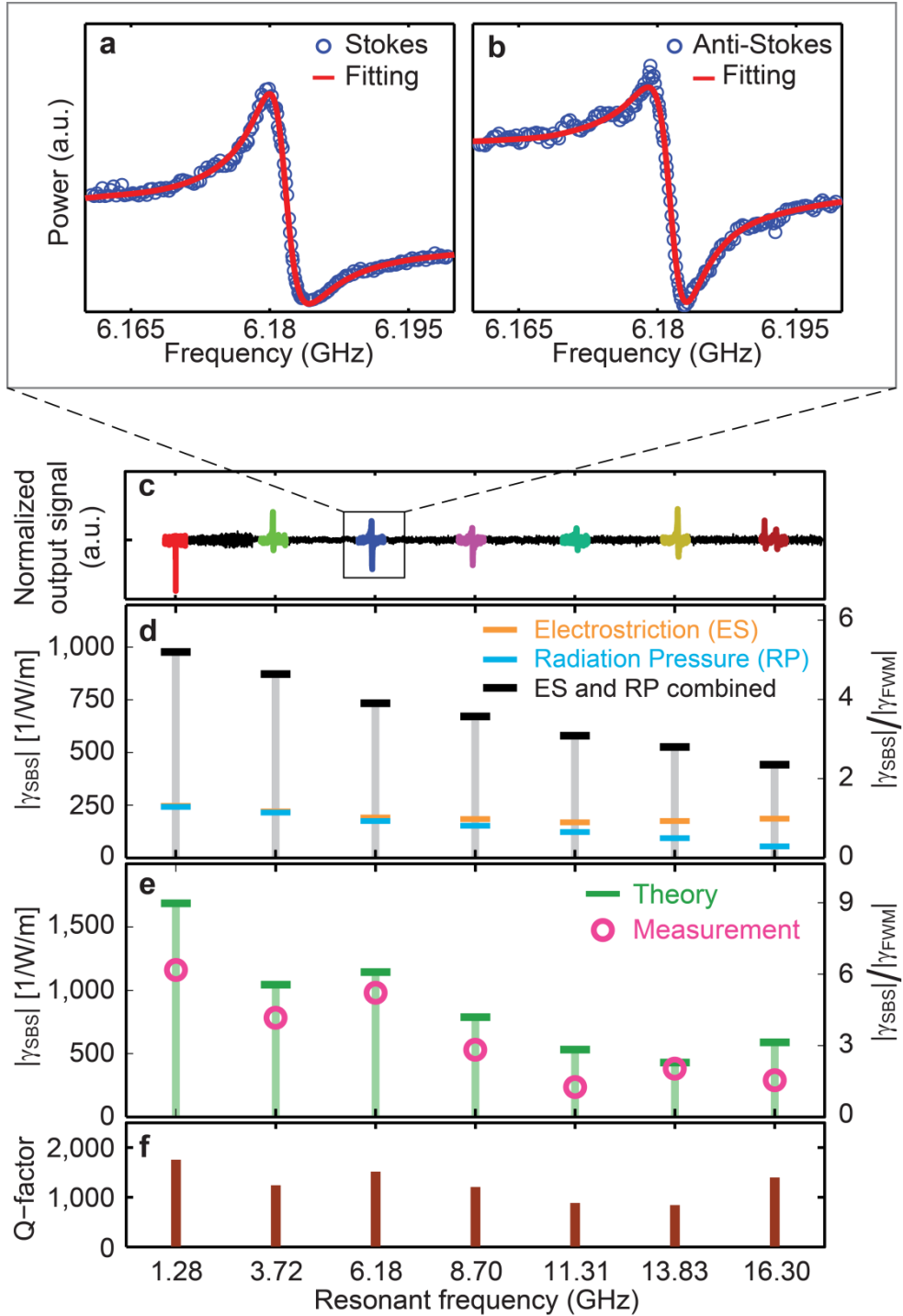

**Figure 5 Characteristic spectral Brillouin line-shapes and nonlinear coefficients obtained from both experiments and simulations. a** and **b** Spectrally resolved Stokes and anti-Stokes Brillouin lines-shapes respectively for the $m = 3$ resonance of $d = 3.8$ $\mu m$ TOPROW waveguide. The theoretical fit (red line) obtained using the line-shape derived in the supplementary information are shown as red curves atop the experimental data (circles). **c.** Normalized heterodyne signals measured by the RF power meter for $d = 3.8$ $\mu m$. **d.** Simulated contributions of both radiation pressure (blue bar) and electrostrictive forces (orange bar) to the total Brillouin nonlinear coefficient (black bar) are numerically calculated for $d = 3.8$ $\mu m$. **e** Comparison between the experimentally obtained (pink circle) and the theoretically calculated (green bar) SBS gain coefficients when experimentally measured quality factor of each resonant mode is incorporated within simulations. **d.** is the measured Q-factor of each resonant mode versus frequency for each resonance of the $d = 3.8$ $\mu m$ TOPROW waveguide, corresponding to modes $m = 1, 2, ... 7$.

Based on the coupled amplitude model described in the Supplement, the magnitude of the Brillouin nonlinear coefficient, $\boldsymbol{\gamma_{SBS}}$, is extracted from the experimental line-shape of both the Stokes and anti-Stokes signatures of each resonance of the $d = 3.8 \, \mu m$ TOPROW waveguide. Seven resonances, spanning frequencies from 1.28-16.30 GHz, are shown in Fig. 5c. The peak value of $|\gamma_{SBS}|/|\gamma_{FWM}|$ and the phononic Q-factor of each resonance extracted from experiments (including separately resolved Stokes and anti-Stokes signatures) are shown Fig. 5e and Fig. 5f respectively. The peak value of the Brillouin nonlinear coefficient at 1.28 GHz is found to be 6.18 times larger than the Kerr nonlinear coefficient of the waveguide. From the known nonlinearities of silicon[48,49], $|\gamma_{FWM}|$ was precisely computed[50,51] (for further details see the Supplement). Based on the computed values of $|\gamma_{FWM}|$, which were confirmed through experiments, the Brillouin nonlinear coefficient is found to be $|\gamma_{SBS}| \cong 1,164 \ W^{-1}m^{-1}$ over the Brillouin active region of the TOPROW waveguide. Moreover, since the Brillouin nonlinear coefficient is related to the Brillouin gain as $2|\gamma_{SBS}| = G_{SBS}$, this nonlinearity corresponds to a forward-SBS gain of $G_{SBS} \cong 2,328 \ W^{-1}m^{-1}$. Additionally, similar analysis of the $d = 0.8 \, \mu m$ waveguide gives $|\gamma_{SBS}|/|\gamma_{FWM}| = 12.3$, yielding a forward-SBS gain of $G_{SBS} \cong 5,366 \ W^{-1}m^{-1}$. Note that this gain coefficient is more than 3000 times larger than recent demonstrations of forward-SBS in fibres[12], and five times larger than the Raman gain produced by silicon[19,22], making Brillouin nonlinearities the dominant nonlinear mechanism in these silicon waveguides. Despite the fact that this nonlinear response is the aggregate of an ensemble of 26 distinct resonators fabricated along the length of the waveguide, remarkably high mechanical Q-factors ( ~1000) are produced for phonon frequencies from 1.28 to 16.3 GHz.

For comparison with experiments, full-vectorial 3D multi-physics simulations were performed through coupled optical force and elastic wave COMSOL models following the approach outlined in Ref. [34]. The distinct contributions of both electrostrictive forces (orange) and radiation pressure (blue) to the total SBS nonlinear coefficient (black) are shown for each phonon resonance, assuming a fixed mechanical Q-factor of $Q = 1000$, in Fig. 5d. A larger variation in Brillouin nonlinearity is seen from the experimental data (circles of Fig. 5e) than from simulations (red bars of Fig. 5d) due to the variation of the measured phononic Q with frequency (Fig. 5f). However, when the frequency dependence of measured Q-factors is included in simulations (Fig. 5f), excellent agreement between simulations and experiments are obtained over the entire frequency range (Fig. 5e). Hence, these results demonstrate the critical role of both material-induced electrostrictive forces and boundary-induced radiation pressure at nanoscales in determining the strength of Brillouin nonlinearities.

Highly localized electrostrictive and radiation pressure distributions within the waveguide core yield a frequency-dependent Brillouin gain with significant departure from conventional backwards-SBS processes involving bulk acoustic waves. In contrast to the rapid $1/f$ roll-off of Brillouin gain with phonon frequency[34,52,53], the experimental (and simulated) Q-factor normalized Brillouin coefficient varies by less than 40% in magnitude over the entire 1-16 GHz frequency range. Unlike conventional systems where the overlap between the optical force distribution and the phonon mode profile is largely frequency-independent, the complex double-lobed spatial force distributions in the core of the silicon waveguide produce a frequency-dependent overlap with various phonon modes, reshaping the frequency dependence of Brillouin coupling. The effect of spatial force distribution on the frequency dependence of coupling can be clearly seen by comparing the computed contributions of electrostriction and radiation pressure to the Brillouin gain of Fig. 5d. While the radiation pressure contribution diminishes quite rapidly with frequency, the electrostrictive component varies little with frequency. The higher bandwidth of electrostrictive coupling is attributed the higher spatial frequencies of the electrostrictive forces. Consequently, the relatively flat Brillouin gain produces efficient photon-phonon coupling over an unprecedented frequency range.

While slight inhomogeneous broadening due to device dimension variability is apparent from some of the measured Brillouin line-shapes, a remarkably consistent Brillouin gain with high Q-factor resonances is observed across the fabricated wafers. The nearly frequency-independent Q-factors observed through experiments suggest that intrinsic material dissipation is not the dominant phononic loss mechanism[54]. Hence, significant further enhancements in Q-factor and Brillouin gain should be attainable with device refinement.

In conclusion, we have made the first demonstration of travelling-wave Brillouin nonlinearities in silicon photonics through the development of a novel class of hybrid photonic-phononic waveguides. Through quantitative measurements, forward-SBS nonlinear susceptibilities were measured to be 3000 times stronger than any previous waveguide system. Multi-physics simulations reveal that this strong photon-phonon coupling is produced by a constructive combination of electrostrictive forces and radiation pressures at nanoscales. The emergence of large radiation pressure-induced couplings represents a new form of boundary-induced Brillouin nonlinearity[32–34], and a new regime of boundary-mediated Brillouin coupling that arises in sub-wavelength structures. This novel waveguide geometry enables independent control of phononic modes and optomechanical driving forces to yield tailorable Brillouin coupling over exceptionally wide bandwidths. Simultaneous coupling to numerous transverse phonon modes yields a relatively flat Brillouin gain over this entire 1-18 GHz frequency range. The wide-band nature of this photon-phonon coupling results from the highly localized optical forces produced within the nanoscale waveguide. These wide-band and high frequency (18 GHz) characteristics were achieved without the need for ultra-high resolution lithography, significantly extending the frequency range of chip-scale photon-phonon coupling over state-of-the-art cavity optomechanical technologies.

Efficient coupling between a continuum of optical and phononic modes through such chip-scale traveling-wave Brillouin processes opens up a host of wide-band signal processing capabilities with CMOS compatible silicon photonics, including pulse compression[13,42], pulse and waveform synthesis[12,14–16,43], coherent frequency comb generation[12,16,17], variable bandwidth optical amplifiers and filters[43–45], and coherent beam combining schemes[46]. Traveling-wave Brillouin nonlinearities can also produce optical phase conjugation[55] and opto-acoustic isolators[31] that are necessary to reduce signal distortions and eliminate parasitic reflections on silicon chips. Additionally, the highly controllable nature of the phonons emitted by this hybrid photonic-phononic system could enable complementary forms of coherent information transduction through traveling-wave processes analogous to recent cavity optomechanical systems[1,3,56]. Since efficient Brillouin-based photon-phonon conversion is possible over wide bandwidths ($> 20$ GHz), and the Brillouin-emitted phonons can be guided and manipulated on-chip, hybridization of Brillouin device physics with silicon photonics, CMOS, and MEMS could provide a host of new coherent signal processing technologies.

Acknowledgements: Sandia Laboratory is operated by Sandia Co., a Lockheed Martin Company, for the U.S. Department of Energy's NNSA under Contract No. DE-AC04-94AL85000. This work was supported by the DDRE under Air Force Contract No. FA8721-05-C-000, the MesoDynamic Architectures program at DARPA under the direction of Dr. Jeffrey L. Rogers, and Sandia's Laboratory Directed Research and Development program under Dr. Wahid Hermina. We thank Marin Soljačić, Ryan M. Camacho, and Ihab El-Kady for helpful technical discussions involving phononic systems, optomechanics and nonlinear interactions. We thank Jack Harris, Jack Sankey, and Benjamin Zwickl for helpful discussions concerning thermo-elastic dissipation in silicon nitride membranes. We are grateful to Douglas Trotter for guidance through fabrication process development, and Whitney Purvis Rakich for careful reading and critique of this manuscript.

**Methods:**

**Fabrication Methods**

The silicon waveguides were patterned in a silicon-on-insulator with at 3000 nm oxide undercladding using an ASML deep UV scanner, and etched in an AMAT DPS polysilicon etch tool. Following resist strip and standard post-etch and pre-diffusion cleans, LPCVD $Si_3N_4$ of 300 nm thickness was deposited in an SVG series 6000 vertical. A chemical-mechanical polish was used to preferentially thin conformal nitride atop Si waveguide. Hot phosphoric acid etch was used to clear the remaining nitride atop the silicon waveguide. The net result is the waveguide cross-section seen in Fig. 1. The nitride layer was then patterned to form the air-slots seen in Fig. 1. Facet cuts for fibre access were created by patterning thick resist using a 1x mask in a SUSS MA-6 contact aligner and employing deep-RIE etch. The oxide under-cladding was then released in a 49% HF etch.

**Experimental Methods**

The pump beam (at 1556 nm) is modulated using a Mach-Zehnder intensity modulator (IM). Mutually incoherent light from another DFB laser at 1526 nm is used as the probe beam. The probe beam split into two paths to form a heterodyne interferometer. In the upper arm of the interferometer, the probe beam is combined with the pump beam using a wavelength division multiplexer (WDM). Both pump and probe beams are coupled into and out of the waveguide using lensed fibres. The pump wave exiting the device is blocked by an interference filter (IF) such that no pump light could be detected. The probe beam is frequency shifted by $\Delta = -165$ MHz using an acousto-optic modulator (AOM) in the lower arm of the interferometer to form the local oscillator for heterodyne detection. In traversing the Brillouin device in the upper arm of the interferometer, the pump-beam produces nonlinearly induces side-bands (or a signal) on the probe-beam. At the output of the interferometer, local oscillator is then mixed with the probe beam and signal using a 30:70 directional coupler, and detected using a high speed (18GHz) receiver. The RF signals produced through detection are then measured with either an RF power meter or an RF spectrum analyser. Due to the -165 MHz frequency offset of the local oscillator, the Stokes and anti-Stokes signatures can be observed as separate tones in the RF spectrum analyser, with a total frequency separation of 330 MHz. By scanning the drive frequency of the RF signal generator (RFG) while measuring either RF power or RF spectrum, detailed analysis of the Brillouin response of the system is made. A fibre-to-chip coupling loss of 8 dB, and waveguide propagation loss of 7 dB/cm were estimated through waveguide cutback measurements. Pump and probe powers internal to the waveguide are estimated to be 6.5 mW and 9.6 mW respectively.

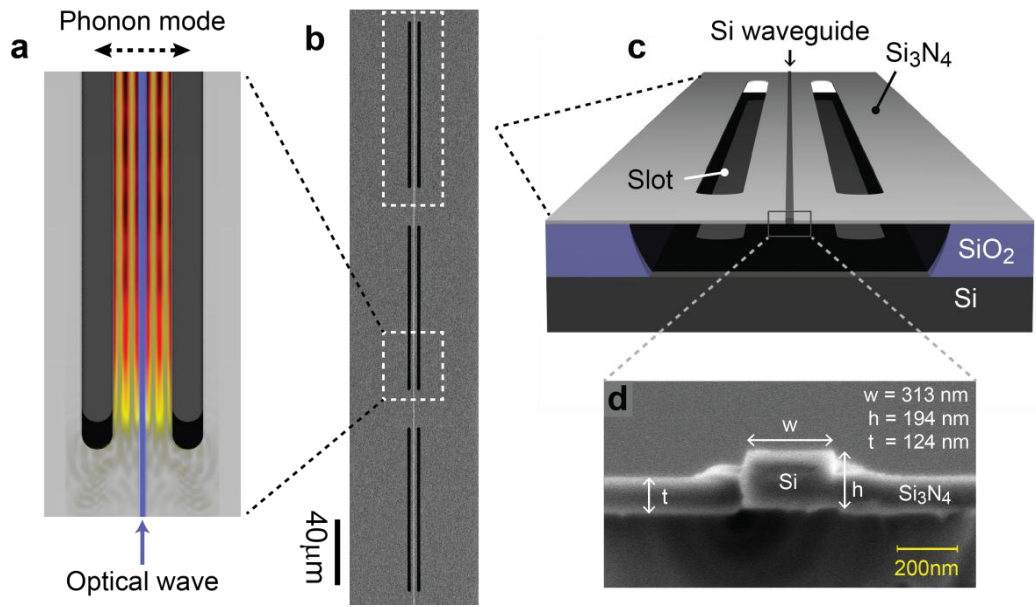

**a** Phonon mode

Optical wave

**b** 40μm

**c** Si waveguide    Si₃N₄

Slot

SiO₂

Si

**d** w = 313 nm
h = 194 nm
t = 124 nm

w

Si    h    Si₃N₄

t

200nm

**Figure 1 Structure of fabricated transverse phonon-resonator waveguide. a** Top-down view of a segment of the Brillouin-active waveguide with optical and phonon modes superimposed. Phonon mode displacement is displayed with a red-yellow colour map, while the guided optical mode is in blue. **b** Top-down SEM image of a segment of the transverse phonon-resonator waveguide (3 phononic resonators shown). **c** Schematic showing the anatomy of the hybrid photonic-phononic waveguide system (1 phononic resonator shown). **d** High resolution SEM cross-section of the silicon waveguide core within the nitride membrane.

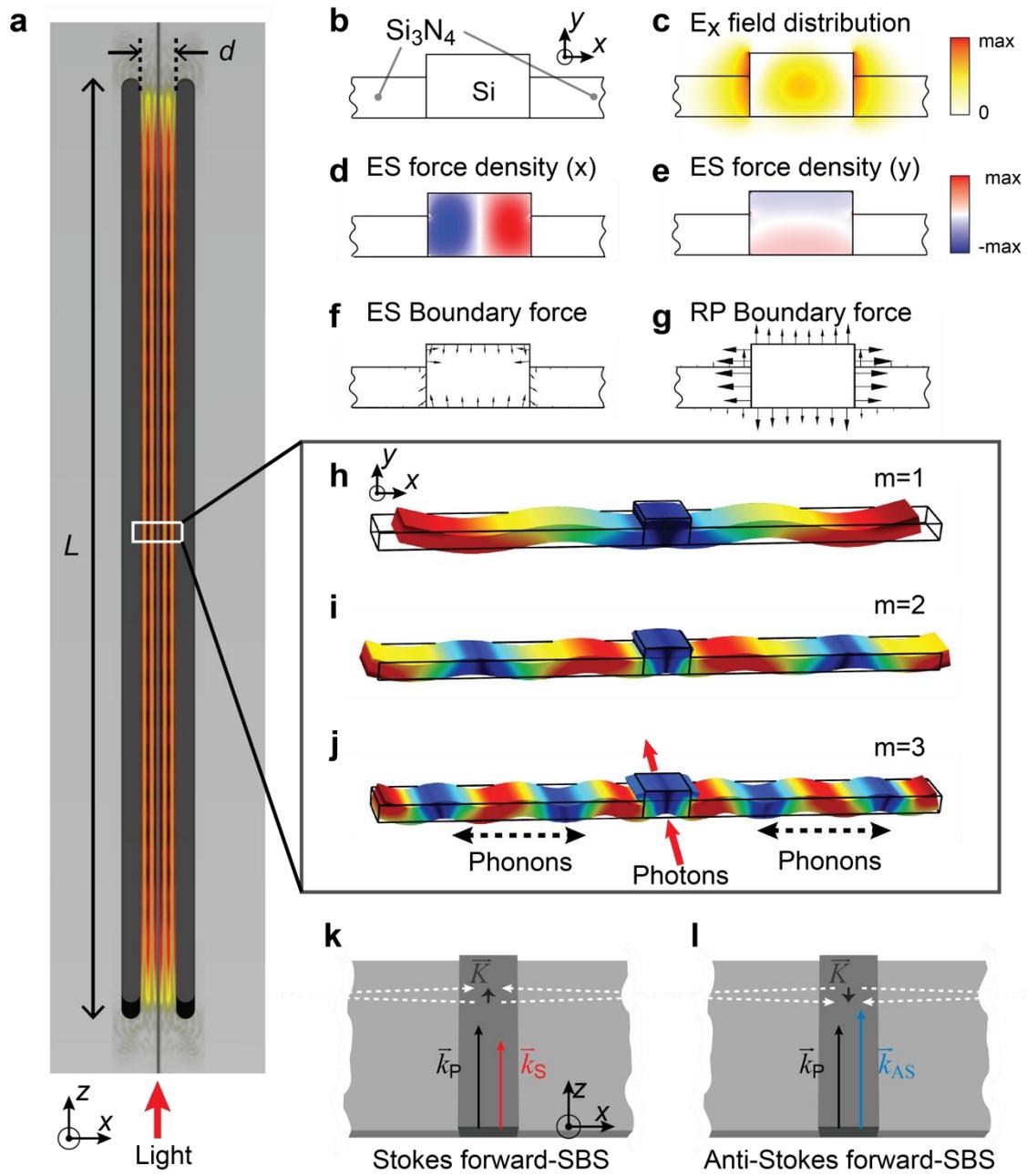

**a**

**b** Si₃N₄  Si  y x

**c** Eₓ field distribution  max  0

**d** ES force density (x)

**e** ES force density (y)  max  -max

**f** ES Boundary force

**g** RP Boundary force

**h** y x  m=1

**i** m=2

**j** m=3

Phonons  Photons  Phonons

**k** Stokes forward-SBS

**l** Anti-Stokes forward-SBS

Light

**Figure 2 Simulations of the transversely oriented phonon-resonator waveguide a** The displacement field of a 3.7 GHz extended phonon excited by optical forces within a TOPROW waveguide with $d = 3.8\,\mu m$ and $L = 100\,\mu m$.  **b-c** The waveguide cross-section and the computed $\boldsymbol{E_x}$ field profile of the optical mode. **d-g** Computed force distributions associated with the guided optical mode. **d** and **e** $x-$ and $y-$ components of the electrostrictive (ES) force densities generated within silicon, respectively. **f** and **g** Electrostrictive and radiation pressure (RP) induced boundary forces produced by the optical mode. **h-j** Characteristic phonon modes, with symmetric displacement fields about the waveguide core, which are efficiently coupled to the optical force distributions of **d-g**. **k-l** Phase-matching diagrams of the respective Stokes and anti-Stokes forward-SBS processes. Phase matching necessitates a vanishing longitudinal phonon wave-vector through forward-SBS, resulting in standing phonon modes (or slow group velocity resonant guided phononic modes) with large transverse wave-vector.

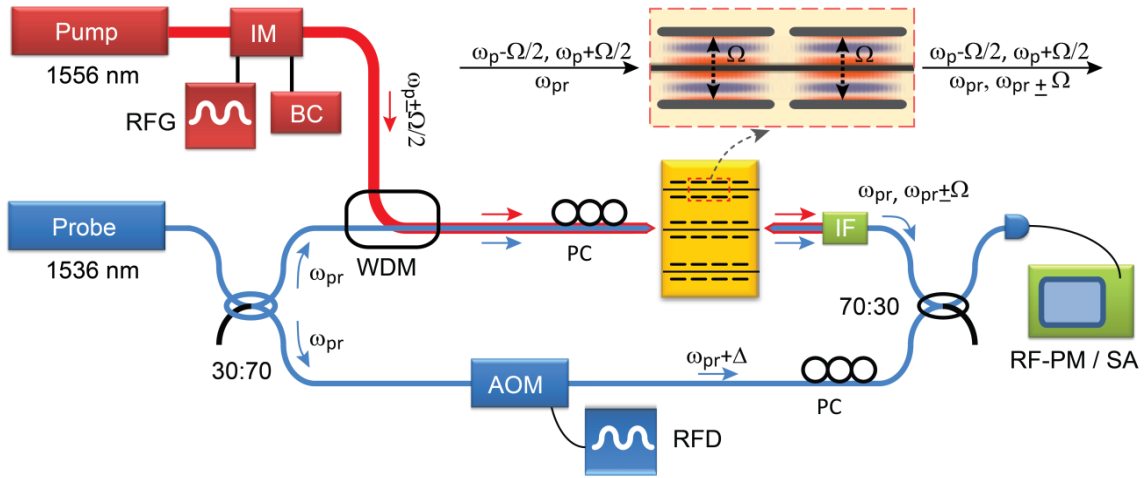

**Figure 3 Heterodyne two-colour pump-probe apparatus used to measure the Brillouin nonlinearity of the TOPROW waveguide.** The nonlinearly induced phase changes imparted to the probe beam (1526 nm) by the waveguide device are detected using heterodyne interferometer, while a modulated pump beam (1556 nm) produces driven excitation of the Brillouin active phonon modes. Components of the apparatus are labelled as follows: ISO is isolator; PC is polarization controller; IM is intensity modulator; RFG is RF generator; BC is bias controller; WDM is wavelength division multiplexer; AOM is acousto-optic modulator; RFD is RF driver; IF is interference filter; SA is spectrum analyser; RF-PM is RF power meter.

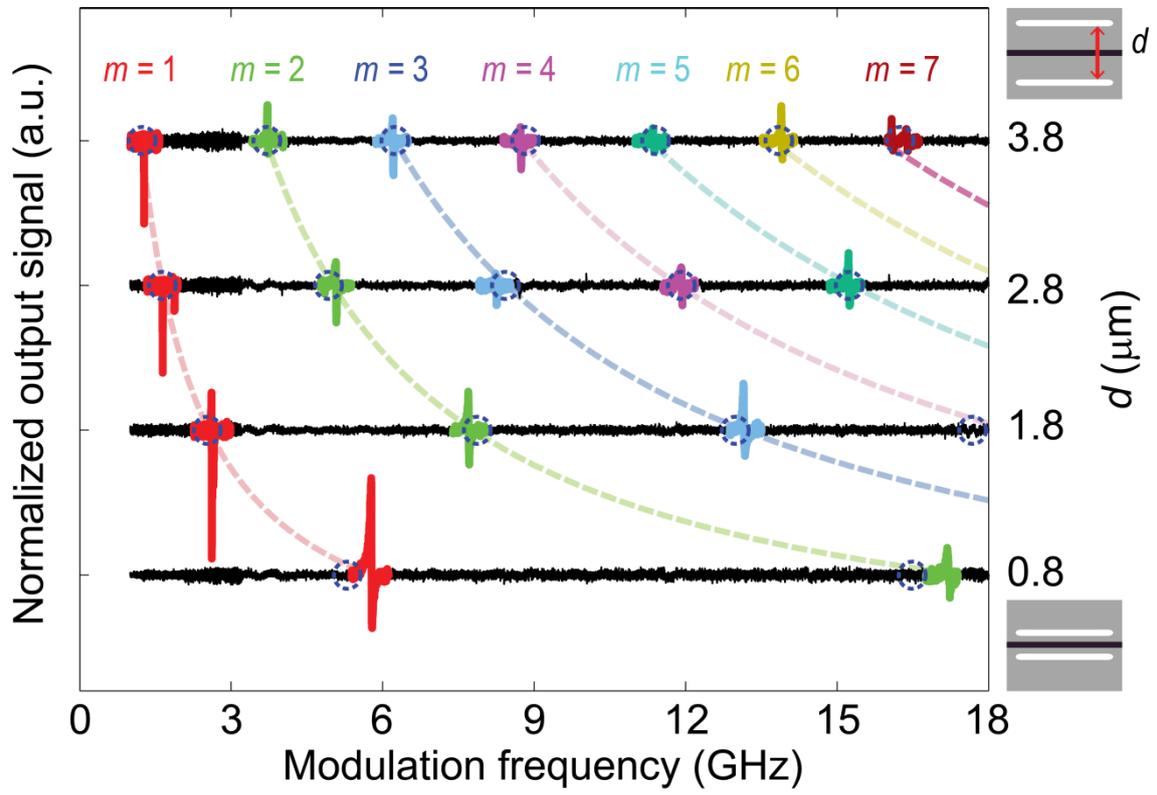

**Figure 4 Spectra of nonlinear Brillouin spectra obtained through heterodyne four-wave mixing measurements.** The resonant Brillouin signatures produced by several TOPROW waveguides are displayed for resonator dimensions of $d = [0.8, 1.8, 2.8, 3.8]\,\mu m$. Each trace is obtained by normalizing the signal produced by each Brillouin-active waveguide to that of an identical reference silicon waveguide (one that is not Brillouin active) under identical experimental conditions. The dashed blue circles shown atop the experimental traces show the numerically computed frequencies of each Brillouin-active phonon mode as the resonator geometry varies. The upper right inset is a schematic the resonator geometry. Each waveguide has the resonant modes at frequencies of $\omega_0(2m - 1)$, where $m = 1,2,3, \dots$ is the order of the resonant modes and $\omega_0$ is the fundamental resonant frequency.

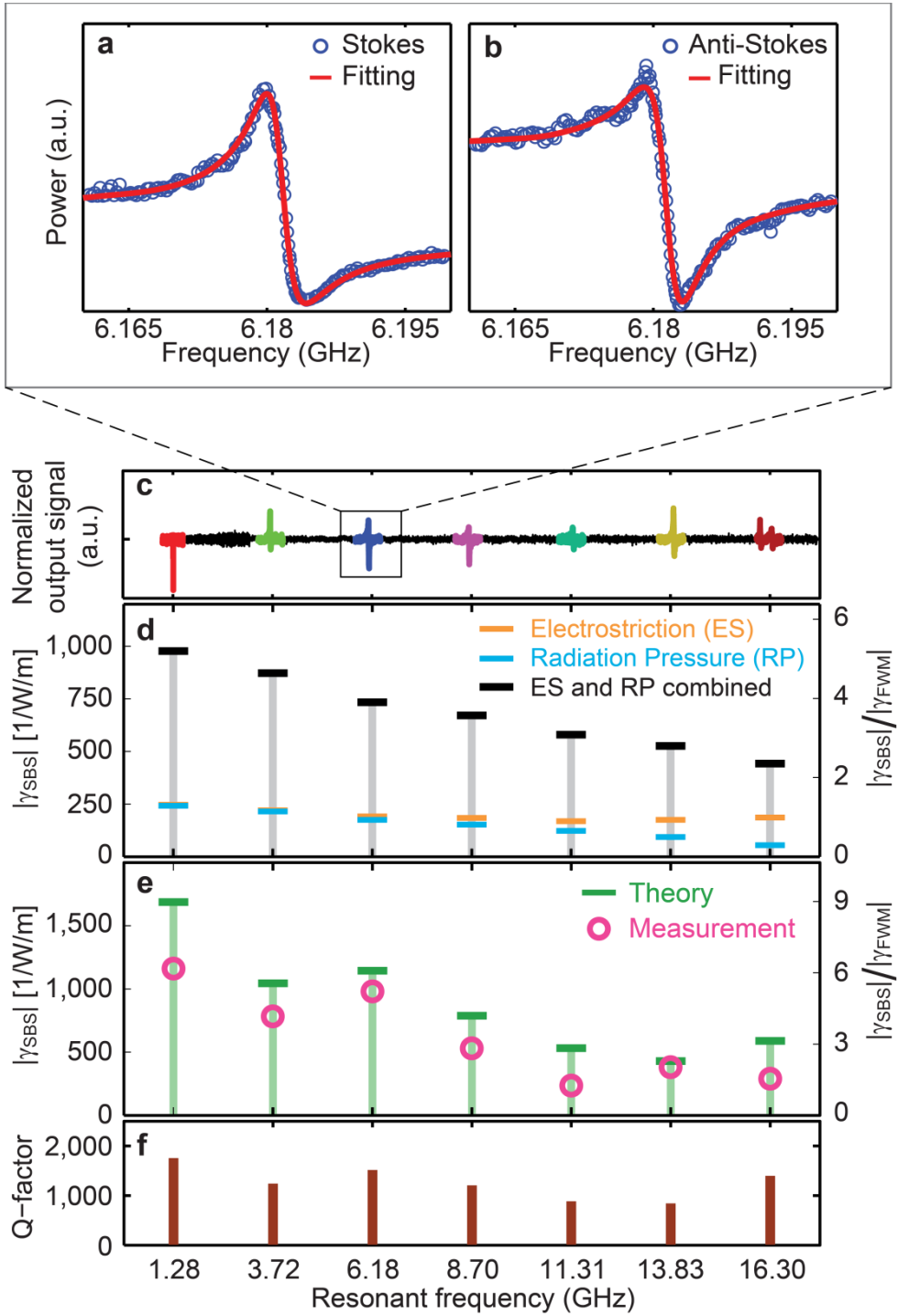

**Figure 5 Characteristic spectral Brillouin line-shapes and nonlinear coefficients obtained from both experiments and simulations. a** and **b** Spectrally resolved Stokes and anti-Stokes Brillouin lines-shapes respectively for the $m = 3$ resonance of $d = 3.8\,\mu m$ TOPROW waveguide. The theoretical fit (red line) obtained using the line-shape derived in the supplement are shown as red curves atop the experimental data (circles). **c.** Normalized heterodyne signals measured by the RF power meter for $d = 3.8\,\mu m$. **d.** Simulated contributions of both radiation pressure (blue bar) and electrostrictive forces (orange bar) to the total Brillouin nonlinear coefficient (black bar) are numerically calculated for $d = 3.8\,\mu m$. **e** Comparison between the experimentally obtained (pink circle) and the theoretically calculated (green bar) SBS gain coefficients when experimentally measured quality factor of each resonant mode is incorporated within simulations. **d.** is the measured Q-factor of each resonant mode versus frequency for each resonance of the $d = 3.8\,\mu m$ TOPROW waveguide, corresponding to modes $m = 1, 2, \ldots 7$.

**Supplementary Information**

In this supplement, we develop the coupled wave equations which describe the nonlinear wave-mixing processes within our Brillouin waveguides, and derive functional form of the asymmetric line-shapes observed through heterodyne pump-probe experiments. Using the analytically derived line-shapes, quantitative analyses of the experimental signatures are performed to determine the magnitude of the Brillouin nonlinear coefficient.

Through experimental arrangement described in the body of this manuscript mutually incoherent pump and probe beams are coupled into the Brillouin waveguide. The pump beam is produced by intensity modulation of the monochromatic laser line. Modulation at frequency $\Omega$, generates a pump beam consisting of two frequencies $\omega_1$ and $\omega_2$ with corresponding wave-amplitudes $A_1$ and $A_2$, where $\omega_2 - \omega_1 = \Omega$. The probe beam consists of a monochromatic wave of disparate wavelength, with wave amplitude $A_3$ and frequency $\omega_3$. Nonlinear wave-mixing processes involving $A_1, A_2$ and $A_3$ generate Stokes and anti-Stokes fields at frequencies $\omega_s = \omega_3 - \Omega$, and $\omega_a = \omega_3 + \Omega$, with wave amplitudes $A_s$ and $A_a$ respectively. These $A_s$ and $A_a$ wave-amplitudes are measured through heterodyne detection to produce the line-shapes discussed in the body of this manuscript. For simplicity, we assume that both pump and probe waves are coupled to TE-like waveguide mode. Since the Stokes and anti-Stokes waves have zero amplitude at the waveguide entrance, the coupled wave equations for Stokes and anti-Stokes wave growth can, to first order, be expressed as [S1,S2,S3] ,

$$\frac{dA_s}{dz} = i\left[\gamma_{\mathrm{SBS}}^{(3)^*}(\Omega) + 2\gamma_{\mathrm{FWM}}^{(3)} + \gamma_{\mathrm{FC}}^{(5)}(-\Omega)P_0\right]A_1^* A_2 A_3 \tag{S1a}$$

$$\frac{dA_a}{dz} = i\left[\gamma_{\mathrm{SBS}}^{(3)}(\Omega) + 2\gamma_{\mathrm{FWM}}^{(3)} + \gamma_{\mathrm{FC}}^{(5)}(+\Omega)P_0\right]A_1 A_2^* A_3 \ . \tag{S1b}$$

Here, $P_0 = 2(|A_1|^2 + |A_2|^2 + |A_3|^2)$, and $\gamma_{\mathrm{SBS}}^{(3)}(\Omega)$ and $\gamma_{\mathrm{FWM}}^{(3)}$ are the third order nonlinear coefficients for Brillouin scattering (SBS) and non-degenerate four-wave mixing (FWM), respectively. Above, we have also neglected two-photon absorption (TPA) induced attenuation of $A_s$ and $A_p$, since in this small signal limit, these terms are much smaller than the source terms of Eq. S1a and Eq. S1b. We assume that the Brillouin nonlinearity, $\gamma_{\text{SBS}}^{(3)}(\Omega)$, is described by a single oscillator, yielding a Lorentzian line-shape of the form [S1],

$$\gamma_{\text{SBS}}^{(3)}(\Omega) = \frac{G}{2} \frac{\Omega_m/2Q}{\Omega_m - \Omega - i\,\Omega_m/2Q}. \tag{S2}$$

Above, $\Omega_m$ is the resonant frequency of the $m^{th}$ mode, $Q$ indicates the quality factor of the phonon resonator, and $G = 2\left|\gamma_{\text{SBS}}^{(3)}(\Omega_m)\right|$ is the Brillouin gain. In addition, $\gamma_{\text{FC}}^{(5)}(\Omega)$ is the fifth order nonlinear coefficient which results from two-photon absorption (TPA) induced the free carrier absorption and refractive index changes imparted by waves $A_1, A_2$ and $A_3$. Solving for time-harmonically modulated TPA-induced free carrier generation rate [S3], and using the carrier rate equation to solve for $\gamma_{\text{FC}}^{(5)}(\Omega)$, one finds,

$$\gamma_{\text{FC}}^{(5)}(\pm\Omega) \equiv -\left(\frac{M}{\tau} \pm \frac{V\Omega}{2}\right)\frac{1}{1/\tau^2 + \Omega^2}\ . \tag{S3}$$

Here $M$ and $V$ are constants with positive value, and $\tau$ is the free carrier lifetime. Note that $\gamma_{\text{FWM}}^{(3)}$ is, to an excellent degree, described as a frequency independent constant which is computed from the waveguide geometry and the nonlinear coefficient of silicon following Refs [S4,S5]. Thus, FWM is non-dispersive, while Brillouin and the free carrier induced nonlinear couplings have frequency dependent responses in our frequency sweeping range. To remain consistent with our experimental arrangement, we note that the FWM and free-carrier effect occur through the waveguide entire waveguide length (4.9 mm), while the Brillouin-active interaction length is shorter than the total waveguide length (2.6 mm). In this case, the optical power of the Stokes field obtained by solving equation (S1a) is given by,

$$g_s = C\left|\gamma_{\text{SBS}}^{(3)^*}(\Omega)L_{\text{SBS}} + \left(2\gamma_{\text{FWM}}^{(3)} + \gamma_{\text{FC}}^{(5)}(-\Omega)P_0\right)L_{\text{tot}}\right|^2 P_1 P_2 P_3\ . \tag{S4}$$

where $C$ is a constant, $P_k$ indicates the optical power of kth field, and $L_{\text{SBS}}$ and $L_{\text{tot}}$ are the interaction lengths of SBS and the rest nonlinear responses, respectively. Equation (S4) consists of two terms, one for Brillouin scattering and another which includes both non-degenerate four-wave mixing and free carrier effects. We refer to the signal from FWM and free carrier effect as the reference signal. In the absence of the Brillouin nonlinearities (e.g. for large detuning from a

Brillouin resonance) the free carrier and FWM contributions to the Stokes sideband can be described as,

$$g_{os} \equiv C L_{tot}^2 \left| 2\gamma_{FWM}^{(3)} + \gamma_{FC}^{(5)}(-\Omega)P_0 \right|^2 P_1 P_2 P_3.$$

(S5)

Since the free carrier effects has slowly varying envelope in frequency, $\gamma_{FC}^{(5)}(\Omega)$ can be treated as a constant in the vicinity of a single Brillouin resonance (e.g. for frequency spans of less than 100 MHz). Since $\gamma_{FC}^{(5)}(-\Omega) \neq \gamma_{FC}^{(5)}(\Omega)$ from equation (S3), the reference signals for Stokes and anti-Stokes are expected to differ from each other when $\Omega$ is comparable with $1/\tau$.

By fitting equation (S4) to the experimentally obtained Stokes and anti-Stokes Brillouin scattering signals as shown in Fig. 5a-b, we can estimate the Brillouin gain $G = 2\left| \gamma_{SBS}^{(3)}(\Omega_m) \right|$. The normalized fitting function $g_s/g_{os}$ is derived from equations (S4-S5) as,

$$\frac{g_s}{g_{os}} = \left| e^{ib_s} + D_n \frac{\Omega_m/2Q}{\Omega_m - \Omega - i\,\Omega_m/2Q} \right|^2,$$

(S6)

where $D_n \equiv G L_{SBS}/\left( 2L_{tot} \left| 2\gamma_{FWM}^{(3)} + \gamma_{FC}^{(5)}(-\Omega)P_0 \right| \right)$ is the relative strength of the Brillouin scattering effect relative to the reference nonlinear responses. Because $\gamma_{SBS}^{(3)}(\Omega)$ and $\gamma_{FC}^{(5)}(\Omega)$ are complex functions, the relative phase between the Brillouin scattering signal and background (FWM + FC) nonlinear responses is defined as $b_s$ in equation (S6). The proportionality to $P_1$, $P_2$, and $P_3$ as well as the constant $C$ in equations (S4) and (S5) are normalized out of equation (S6).

Note that because of the frequency dependent free-carrier effect, different resonant modes are normalized by different nonlinear background. In experiments, we observed that at high frequency (>15 GHz) the amplitude of the reference signal converges to $\left| 2\gamma_{FWM}^{(3)} \right|$, indicating $\left| 2\gamma_{FWM}^{(3)} \right| \gg \left| \gamma_{FC}^{(5)}(\pm\Omega)P_0 \right|$. We can experimentally measure the reference signal spectrum and obtain the ratio $\eta \equiv \left| 2\gamma_{FWM}^{(3)} + \gamma_{FC}^{(5)}(\Omega_m)P_0 \right|/\left| 2\gamma_{FWM}^{(3)} \right|$. Then using established methods for computing $\left| 2\gamma_{FWM}^{(3)} \right|$ based on well-known values for the Kerr nonlinearities of crystalline silicon [S4,S5], we can estimate the Brillouin gain $G$ using the following equation,

$$G = 2D_n\eta\left|2\gamma_{\text{FWM}}^{(3)}\right|\frac{L_{\text{tot}}}{L_{\text{SBS}}}. \tag{S7}$$

Note that we have neglect propagation losses in S1a and S1b, as losses do not alter the final functional form of the derived line-shape in the small signal limit.

The magnitude of $\gamma_{\text{FWM}}^{(3)}$ produced by the silicon waveguide was computed using the accepted Kerr coefficient of $n_2 = 4.5 \cdot 10^{-18}\ [m^2/W\ ]$ in silicon [S3]. Employing the full-vectorial method for computing $\gamma_{\text{FWM}}^{(3)}$ described in Ref. [S4], $\left|\gamma_{\text{FWM}}^{(3)}\right|$ was computed to be $188\ [1/W/m]$ for TOPROW waveguides with nitride membrane widths of $d = [1.8,\ 2.8,\ 3.8]\ \mu m$. As the nitride width was reduced to $d = 0.8\ \mu m$, the close proximity of the lateral nitride boundary increases the modal confinement, yielding $\left|\gamma_{\text{FWM}}^{(3)}\right|$ of $214\ [1/W/m]$.